\documentclass[%
 reprint,%
secnumarabic,%
amssymb, amsmath,%
aip,cha,%
groupedaddress,%
frontmatterverbose,
]{revtex4-1}

\usepackage{graphicx}%
\usepackage{dcolumn}%
\usepackage{docs}%
\usepackage{bm}%
\usepackage[colorlinks=true,linkcolor=blue]{hyperref}%
\usepackage{float}

\expandafter\ifx\csname package@font\endcsname\relax\else
 \expandafter\expandafter
 \expandafter\usepackage
 \expandafter\expandafter
 \expandafter{\csname package@font\endcsname}%
\fi
\begin{document}

\title{Complex network analysis of water distribution systems}%

\author{Alireza Yazdani}
\author{Paul Jeffrey}
\email{p.j.jeffrey@cranfield.ac.uk}
\affiliation{School of Applied Sciences, Cranfield University, MK43 0AL, UK}%

\begin{abstract}
This paper explores a variety of strategies for understanding the formation, structure, efficiency and vulnerability of water distribution networks. Water supply systems are studied as spatially organized networks for which the practical applications of abstract evaluation methods are critically evaluated. Empirical data from benchmark networks are used to study the interplay between network structure and operational efficiency, reliability and robustness. Structural measurements are undertaken to quantify properties such as redundancy and optimal-connectivity, herein proposed as constraints in network design optimization problems. The role of the supply-demand structure towards system efficiency is studied and an assessment of the vulnerability to failures based on the disconnection of nodes from the source(s) is undertaken. The absence of conventional degree-based hubs (observed through uncorrelated non-heterogeneous sparse topologies) prompts an alternative approach to studying structural vulnerability based on the identification of network cut-sets and optimal connectivity invariants. A discussion on the scope, limitations and possible future directions of this research is provided.
\end{abstract}

\maketitle

\date{\today}%
\revised{\today}%

\begin{quotation}
Modern society is highly dependent on the reliable performance of critical infrastructures such as water, energy and transport systems. The increasing level of complexity and interdependence of such systems poses numerous challenges to reliable design and optimal control, hence the need for such issues to be supported with insights generated beyond the traditional engineering disciplines. This provides an opportunity for complex networks researchers to apply new concepts and tools to describe, predict and control the behavior and evolution of critical infrastructure systems. Water distribution systems, one of the most important complex infrastructure systems, can be represented as networks of multiple interconnected interacting parts. This work is a study of the structure, connectivity and building blocks of the networks underlying such systems and identifies the relationship between the structure of water distribution systems and their operational reliability and susceptibility to damage. Some important similarities and differences between water supply systems and other complex infrastructure networks are explained and the role of the supply-demand structure in the formation and operation of water distribution systems has been highlighted.
 \end{quotation}

\section{Introduction}
\label{sec:1}
Research in the field of complex networks and their structural properties has grown rapidly in the past few years \cite{1,2,3,4,5,6}. Complex networks are usually understood as distributed systems consisting of multiple interconnected components structured in non-trivial configurations in which the network function is largely affected by the structure \cite{7}, depending on the organizational complexity and the level of interaction among the components. One significant demonstration of such interplay of the network structure and function is observed in the study of so-called scale-free networks \cite{2}, characterized by heterogeneous structures and non-uniform degree distributions in which the great majority of the nodes have very low connectivity and a few nodes, known as hubs, are highly connected. Scale-free networks reveal important properties in terms of the level of resilience (or lack of it) when exposed to errors and attacks; they are robust against random failures but vulnerable to targeted attacks on their hubs \cite{8}.\\ 

The ubiquity and importance of complex networks observed as the underlying structural framework of many technological, information and social systems has urged researchers to study the dynamics of network formation and growth, which in turn has given rise to efforts to understand the structural vulnerability of networks and their resilience against perturbations, random failures and targeted attacks \cite{8,9,10,12,13}. Due to the increasing level of complexity and component interdependency in critical infrastructure networks, several studies have focused on understanding the security of these networks and their susceptibility to damage. Examples of infrastructure networks include urban roads, rail network, power grid, gas pipeline networks, water distribution networks and supply chains \cite{9, 10,14,15,16,17}. By construction, most of these networks are spatially organized planar graphs. Such a property imposes severe limitations on network connectivity and layout, and hence they are studied differently from other non-technological complex networks \cite{19}.\\

Water distribution networks (WDNs) are among such spatially organized systems in which multiple assets are connected by actual physical links. In a link-node representation of physical components in water distribution networks, pipes and other connections are shown by edges, with the fixed junctions (reservoirs, tanks and demand points) and pipe intersections represented by nodes. WDNs are complex in the sense that their multiple interconnected components are arranged in non-trivial configurations and interact in complex ways. Some important contributors to the complexity and uncertainty in design and operation of WDNs are: the range of possible combinations of the pipe sizes, materials and connectivity layouts, location of the valves and pumping stations, capacity of tanks, control valve settings, pump scheduling and unknown demand for water.\\

The management of WDNs depends on system layout (topology and patterns of connectivity), design (system sizing) and system operation (given a design) \cite{20}. The optimal design of large water distribution networks is a complex problem that involves making decisions on pipe layout and sizes (length and diameter), while trying to minimize the cost of network design, building and operation. This problem can be formulated as the problem of minimizing costs subject to hydraulic feasibility, satisfaction of demands and meeting pressure constraints \cite{21}. Consequently, numerous quantitative and simulation methods on the least-cost design of water distribution systems have been developed (see \cite{ 21, 22} and references therein). Depending on the size and complexity of the design problem, these methods employ techniques such as: linear programming \cite{23}, non-linear programming \cite{24}, integer goal programming \cite{25}, Monte Carlo simulation \cite{26} and evolutionary methods including genetic algorithms and ant colony optimization \cite{27,28,29}.\\

In addition to the technical and computational complexities, there exist important issues during the design of WDNs such as redundancy (the existence of alternative resources or supply paths) and reliability (the probability of non-failure over time). The assessment of water distribution systems reliability is a daunting task that largely depends on availability of historical data for mechanical component failures and hydraulic failures. Fortunately, water distribution system reliability is largely defined by its network layout (e.g. redundancy improves reliability) \cite{30}. However, regardless of the utilized method, optimization reduces cost by reducing pipe diameter (i.e. reducing the capacity) or by completely eliminating the link between nodes (i.e. eliminating the loops in the network) and hence reducing redundancy. This, in the absence of suitable optimization constraints for redundancy and optimal connectivity, makes the system unreliable and largely vulnerable to the failures of links and nodes following errors, attacks and overloads. Consequently, the analysis of network topology and measurement of the redundancy and optimal connectivity, to be used in the framework of optimization design models, could make a significant contribution to this field.\\ 

It is also worth noting that, with more attention being paid to the development of protective measures to increase network invulnerability and robustness during the strategic planning of infrastructure networks, deterministic methods based on graph invariants or complex networks can be employed to compare alternative designs and assess network efficiency and overall robustness against failures. Moreover, given recent developments in the analysis of the structure of technological and infrastructure networks \cite{10, 12,14,15,19} and the similarity among spatially organized complex systems, the study of the structural vulnerability of WDNs seems timely and relevant. To the authors' knowledge, the application of complex networks approaches in water supply systems is limited to a study of the small world phenomena in WDNs \cite{31}, with no systematic study of WDN structure and function reported by using complex networks methodology. This study intends to project the important findings of network-based approaches to the analysis of technological systems onto WDNs and thereby prompt a dialog between theoretical network scientists on one hand, and engineers and operational researchers on the other hand. Such a dialog is central to improving our ability to overcome the numerous challenges encountered in the design, operation and protection of complex infrastructure systems.\\

In this paper, the structural organization of water distribution systems are studied and compared to similar types of physical networks reported in the literature. The assessment is achieved through an analysis of empirical data for four benchmark water distribution networks based on measurements that quantify the structure of the network paths, cycles, connectivity and efficiency. Moreover, network resilience is studied by examining two important topological features; robustness and path-redundancy. Robustness is viewed as the overall structural tolerance to errors and failures, and redundancy as the existence of alternative supply paths, usually observed in the form of the loops or equivalent structures consisting of nodes and standby links or components not employed to their full capacity. Network robustness is analyzed by examining the spectra of the connectivity and Laplacian matrices of the studied WDNs and, particularly, by evaluating the two descriptive graph invariants of algebraic connectivity and spectral gap, which quantify static fault tolerance and optimal-connectivity, respectively. Based on such a viewpoint, the structural vulnerability of representative WDNs is revealed in terms of the presence of the bridges and cut sets. Finally, a discussion on the scope and the limitations of the presented methodology and possible future directions of this research is presented.
\begin{figure*}[t]
\centering
\includegraphics[width=0.8\textwidth, height=9.80cm]{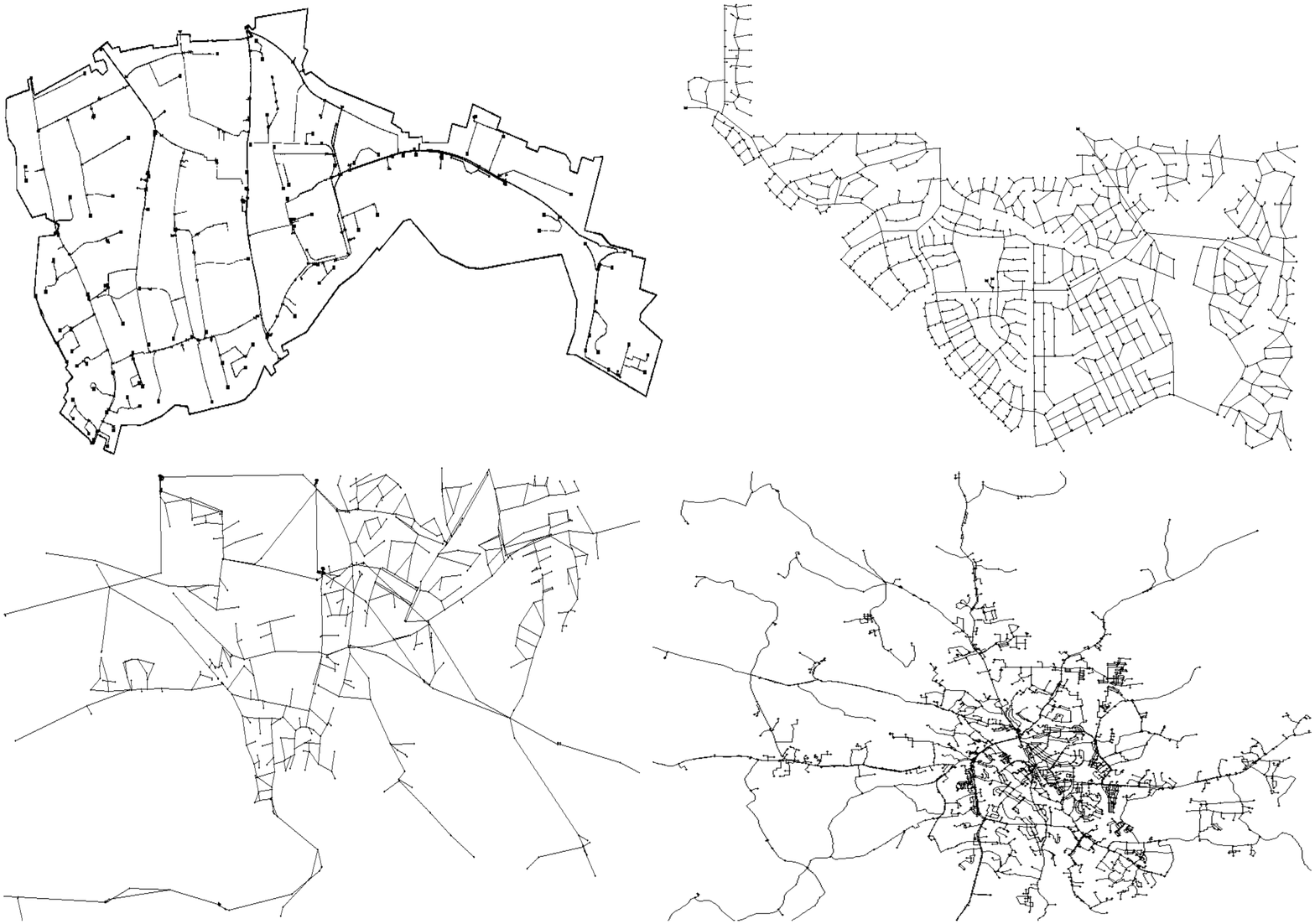}
\caption{The graph representation of the studied water distribution networks; East-Mersea (top-left), Colorado Springs (top-right), Richmond (bottom-left), Kumasi (bottom-right).}
\label{fig:1}
\end{figure*}

\section{Water distribution networks}
\label{sec:2}
In general, the physical configuration of water distribution networks is dependent on the geographical organization of the supply and demand nodes and the location of physical barriers such as roads, buildings, rivers and so on. The nodes in a WDN are typically grouped by sources (e.g. reservoirs, tanks and storage facilities), control and distribution nodes (e.g. pressure control valves, pipe junctions, pumps) and demand nodes or sinks (e.g. consumers). WDN links, on the other hand are capacitated transmission and distribution pipes with specified length, size and other physical attributes. However, the direction of the links in WDNs is subject to occasional changes (except for the pipes attached to a source or a sink) due to operational flow and pressure requirements, considerations related to the pumping cost and flow redirections that might take place following the failure of a major supply path. \\

A comprehensive assessment of WDN resilience should take into account the non-topological specifications of the network components including the size of the links and importance and influence of the nodes. Such an approach will enable the establishment of realistic correlations between the topology of the network and those operational aspects important to the analysis of reliability and vulnerability, such as the total amount of lost water and the equivalent hours of disruption as a result of failures. Achieving this, however, largely depends on the study of the flow dynamics throughout the network and the analysis of empirical pressure and flow data (not available here) followed by extensive simulations (computationally costly) in order to establish and validate the described correlations. \\

Therefore, in this contribution the studied networks are treated as undirected graphs, and a simplified approach is adopted which is based on the statistical properties of network topology and applications of graph theory to identify the structural patterns and building blocks of the networks. Such an approach provides a conceptual study framework and establishes some necessary but perhaps insufficient conditions to fully assess network vulnerability. Meanwhile, it is worth mentioning that these networks are normally regarded as connected graphs in the sense that, under normal operational circumstances, there exists at least one path between every two nodes and in particular between the water supply sources (e.g. reservoirs, tanks) and the consumers. Consequently, great emphasis has been placed here on graph connectivity and the ways to preserve and improve it, as any disconnection between the source and the consumers is regarded as a failure.\\

In this work, four benchmark real distribution networks are studied (see Fig. 1), each representing different formation and organizational patterns. “East-Mersea” is a small distribution sub-network owned by Anglian Water Services in the UK. The “Colorado Springs” network reported in \cite{32} and the “Kumasi” town water distribution network in Ghana in Africa are examples of networks with multiple water supply sources. Finally, the “Richmond” network is a sub-network of the Yorkshire Water system in the UK with one single reservoir as reported in \cite{33}. Due to the technical difficulties and high expenses associated with obtaining the data on WDN components that are located underground, datasets on WDNs are not widely available for this type of analysis and consequently the studied datasets represent a small sample of the existing set of water supply systems.\\

As is common to the representation of WDNs, reservoirs, tanks, control valves, pipe junctions, pumps and demand nodes are represented by nodes and transmission and distribution pipes are regarded as graph links. The studied networks seem to have formed during uncontrolled gradual expansion over time. While “Colorado Springs” is significantly more looped-like than “Richmond” and its structure is somewhat ordered as a lattice, at least locally, it cannot be definitely stated whether its global ordering has been obtained as result of a single optimized construction plan. The irregularity in structure is much more visible in the “Richmond” example where the network layout largely deviates from the lattices. This structural property of WDNs may be interpreted as local robustness at the expense of global robustness \cite{34}.
\section{Structural measurements}
\label{sec:3}
Each network is modeled as a mathematical graph $G=G(N,E)$ in which $N$ is the set of $n$ graph nodes and $E$ is the set of $m$ graph edges. Link density for an undirected network is given by $q=\frac{2m}{n(n-1)}$ , the fraction between the total and the maximum possible links to indicate the sparseness or dense-connectivity of network layout. The four studied WDNs are sparse, in the sense that the number of graph links is far from maximal, as observed by low link density values (Table I). A graph is planar if it can be embedded in the plane so that its edges intersect only at a node mutually incident with them. While it may not be possible to prove the strict planarity of the studied WDNs (for example by using Kuratowski's characterization theorem \cite{35}), only a negligible percentage of the edge intersections do not match to their endpoints and hence the studied networks are near-planar, similar to other spatially organized infrastructure networks \cite{12,15,19}. This is not surprising given the fact that in the design of water distribution systems, it is not very common or even feasible to lay transmission or distribution pipes (elevated or otherwise) in multiple layers on top of each other and hence a typical water distribution network is usually organized in a single-layer almost planar structure.\\

One simple way to determine the overall similarity to or alternatively the deviation of the network structures from tree-like or mesh graphs is to evaluate the link-per-node ratio $e=\frac{m}{n}$ or the average number of connections per node or mean node-degree $<k>$ related by the equation $e=\frac{m}{n}$. The link per node ratios for spatially organized networks including the studied WDNs lay between the two limits of $e=1$ and $e=2$ which represent tree-like planar graphs and two-dimensional (infinite) regular lattices, respectively. In general, the grid-like structures facilitate equalized distribution of flow and pressure under varying demand rates and locations in WDNs, and hence this simple metric may illustrate the hydraulic efficiency of the network to a limited extent. \\

Another metric used to describe the structural organization of WDNs is the central-point dominance $c_b'$ defined in \cite{36}, which, in the analysis of flow networks, may be used to indicate how network flow is controlled by centrally located point(s), or to quantify the degree of concentration of the network layout around a center. Central-point dominance is calculated by taking the mean over the betweenness centrality values of all nodes indexed by the maximum value of betweenness (achieved at the most central point). This is formulated as $c_b'=\frac{1}{n-1}\Sigma_{i}(b_{max}-b_{i})$ where $n$ is the number of nodes, $b_{i}$ is the betweenness centrality of the node $i$ and $b_{max}$ is the maximum betweenness centrality value. The node betweenness centrality is defined \cite{36} as the number of shortest geodesic paths between two given vertices that pass through that node divided by the total number of shortest geodesic paths between those two vertices. Larger values of betweenness centrality indicate that a node (edge) is located on many short paths. It is easily verified that $c_b'=1$ for wheel or star-like graphs and $c_b'=0$ for regular networks with all points having equal betweenness centralities. Interpretation of the central-point dominance largely depends on a network's specific function and the underlying design considerations. While construction of star-shaped topologies by locating a hub at the center will be more economic and may facilitate transportation in the network, it will significantly compromise network robustness due to the high sensitivity of such network design to the failure of the most central point. In WDN design, highly centralized structures rarely exist, since distributed and grid-like structures are preferred, as discussed earlier.\\
\begin{table*}[]
\centering
\caption{Graph attributes and measurements for the benchmark water networks ($n$ = nodes, $m$ = links, $q$ = link density, $e$ = link per node ratio, $<k>$ = average node-degree, $k_{max}$ = maximum node-degree, $c_b'$= central-point dominance, $c$ = clustering coefficient, $r_m$ = meshed-ness)}
\label{tab:1}     
\begin{ruledtabular}
\begin{tabular}{cccccccccc}
\noalign{\smallskip}
Network   &  $n$    &   $m$  &  $q$   & $e$      & $<k>$ & $k_{max}$    & $c_b'$   & $c$  &  $r_m$\\
\noalign{\smallskip}\noalign{\smallskip}
East-Mersea        &  755     & 769                  &  $2.70 \times 10^{-3} $       & 1.01         &  2.04              & 4                            & $3.6 \times 10^{-1}$              & $0.00$                         & $9.97\times10^{-3}$ \\
Colorado Springs &   1786     & 1994              &  $1.25\times10^{-3} $       & 1.11        &  2.23              & 4                             & $4.2\times10^{-1}$             & $8.82\times10^{-4}$     & $5.86\times10^{-2}$  \\ 
Kumasi                 &  2799	   & 3065	          &  $7.83\times10^{-4} $       & 1.10        &  2.19	             & 4	                    &$ 4.5\times10^{-1}$	        & $1.54\times10^{-2}$     & $4.77\times10^{-2}$ \\
Richmond             &  872	   &  957	         &  $2.52\times10^{-3} $       & 1.09         &  2.19	             & 4	                    &$ 5.6\times10^{-1}$	        & $4.02\times10^{-2}$     & $4.95\times10^{-2}$ \\
\noalign{\smallskip}
\end{tabular}
\end{ruledtabular}
\end{table*}

The metrics introduced so far, only capture very generic information regarding the structure of the studied WDNs which are planar and spatially organized networks with connectivity restrictions. Extending the analysis to include other structural properties such as the structure and number of cycles and loops as an indicator of network redundancy (regarded as an important criterion for reliability and invulnerability in the context of network design) adds an important dimension to the assessment. To this end, the clustering coefficient is a useful measure to characterize the status of network loops of length three, which in this context may be regarded as an indicator of path redundancy and a way to quantify the existence of the looped alternative supply routes which ensure flow between water supply sources and demand points where the direct link or shortest path between these two nodes fails. The network's clustering coefficient is (transitivity) defined by
\begin{equation}
\ c= \frac{3N_\Delta}{N_3}
\label{Equation 1}
\end{equation}
and measures the density of transitive triangles in a network, where $N_\Delta$ is the number of triangles and $N_3$ is the number of connected network triples. However, one major difficulty associated with the use of the clustering coefficient for the study of cycles in spatially organized urban networks is that the dominant looped structures in such networks are non-triangular and mostly quadrilateral \cite{12, 37}. Therefore, the clustering coefficient is not a particularly good indicator of path redundancy in such networks and hence a more general measurement is required to overcome this difficulty. \\

One recently proposed metric to quantify the density of the cycles and loops in planar graphs is known as the meshed-ness coefficient \cite{12} and is more relevant in this respect. In the design of water distribution networks, represented as a graph with $n$ nodes and $m$ edges, the number of independent loops is given by $f=m-n+1$ for single source networks and by $f=m-n$ for multiple source systems \cite{38}, derived from Euler's formula to count the (finite) faces associated with any planar graph. The maximum number of links cannot exceed $3n-6$ in planar graphs. Consequently, the meshed-ness coefficient $r_m$ can be defined as the fraction between the actual number of loops and the maximum possible number of loops (bounded by $2n-5$)
\begin{equation}
\ r_m= \frac{f}{2n-5}
\label{Equation 2}
\end{equation}
which quantifies the density of any kind of loops (not necessarily of triangular) and may be regarded as a surrogate measure of path redundancy in the network. The numerical values of the measurements discussed above as calculated for the studied networks are presented in Table (I).\\
\begin{figure}[]
\centering
\includegraphics[width=0.5\textwidth, height=12cm]{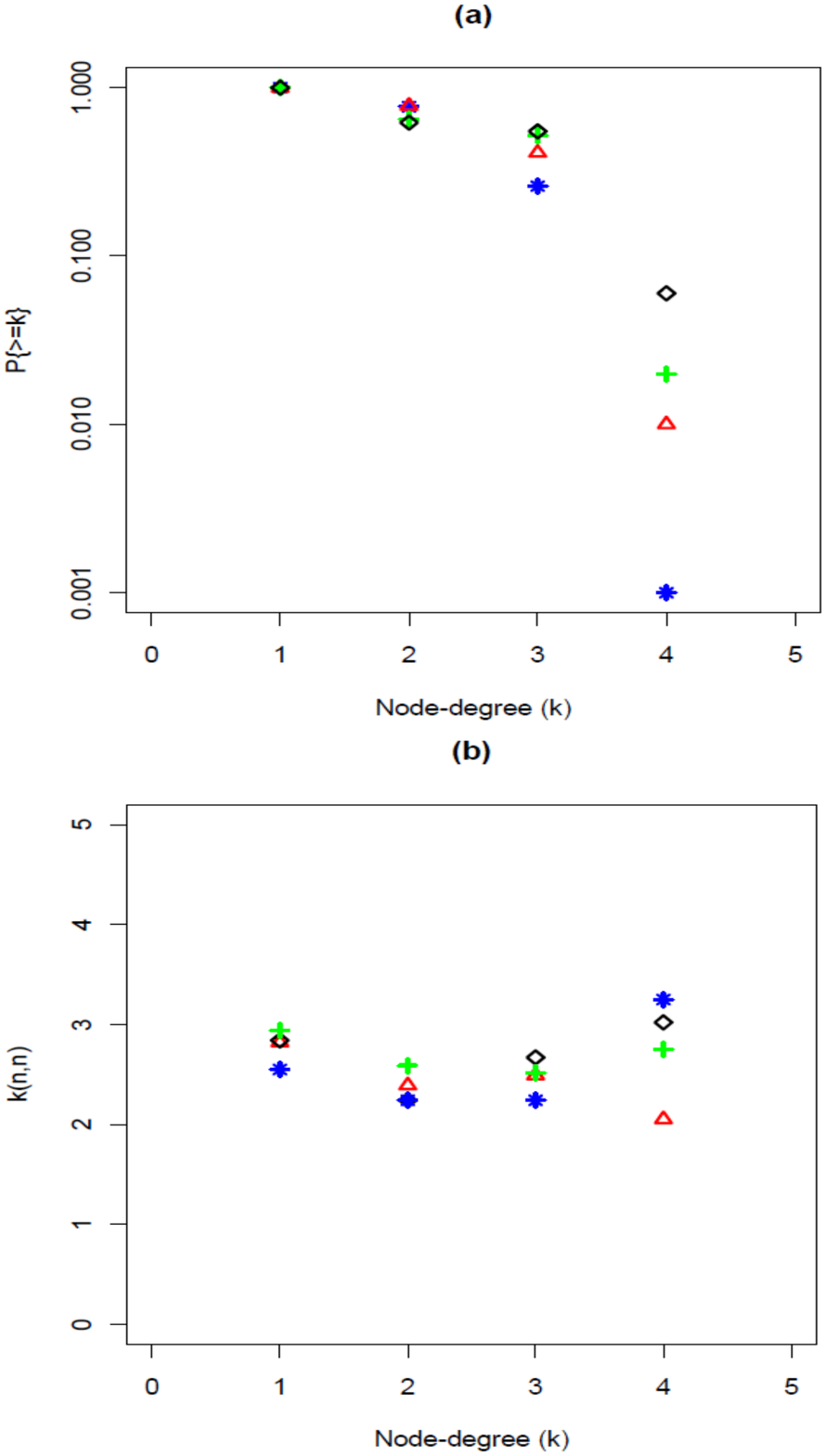}
\caption{Degree distribution and correlation properties of networks; Richmond (red triangles), Colorado Springs (black diamonds), East-Mersea (blue stars) and Kumasi (green crosses) (a) Log-linear cumulative degree distribution plot of the studied networks showing a sharp drop from $k=3$ (b) Nearest neighbor degree correlation $k_{n,n}(k)$.}
\end{figure}

As suggested, planarity and other physical specifications impose severe limitations on the connectivity of WDNs. This is witnessed through the sparseness (low link-density values) of the studied networks and a relatively uniform degree distribution with small differences between the maximum and minimum node degree in these networks. The degree distributions of all four WDNs, illustrated in Fig. 2(a), have a maximum of four connections and a minimum of one connection per node, with the biggest percentage of nodes having degrees three ($48.2$\% for Colorado Springs, $50.7$\% for Kumasi and $39.5$\% for Richmond) and two ($50.6$\% for East-Mersea), respectively. By using the inequality $<k>=\frac{2m}{n} \leq \frac{2(3n-6)}{n}$ for planar graphs, it can be seen that the average node degree is strictly smaller than 6, whereas the average degree distributions obtained in both cases are much lower than this theoretical maximum.\\

The studied examples are found to be single-scaled non-heterogeneous networks (Fig. 2(a)) in which the cumulative probability degree distributions can be approximated by the exponential form $P_{x>k}(k)=\int_x^\infty \! P(k) \, \mathrm{d}k \approx exp(\frac{-k}{\gamma})$. The exponents are found as $\gamma=1.71$ for East-Mersea ($r^2=0.901$), $\gamma=2.10$ for Colorado Springs ($r^2=0.877$), $\gamma=2.01$ for Kumasi ($r^2=0.872$), and $\gamma=1.98$ for Richmond ($r^2=0.892$). These numbers are generally in the same range as those similarly reported for other spatially organized infrastructure networks \cite{12,16,17,39}, in spite of the presumed approximation error due to the curve fitting based on four sample points only. The existence of degree correlation \cite{40} among vertices is calculated by $k_{n,n}(k)=\Sigma_{k'} p(k'\vline k)$ which is the average nearest neighbor degree of a vertex of degree $k$, where $p(k'\vline k)$ is the conditional probability that an edge belonging to a node of degree $k$ points to a node of degree $k$. The relationship between the nodes of a given degree $k$ and the degree of the nearest neighboring nodes, illustrated in Fig. 2(b), does not show any increasing or decreasing trend in any of the four networks. Based on these observations (i.e. relatively uniform degree distribution and the absence of highly connected nodes), no assortative or disassortative connectivity by node degree has been detected and hence the studied networks are observed to be uncorrelated.
\section{Path length and efficiency}
\label{sec:4}
Accessibility is determined by the level of ease or difficulty associated with dispatching a commodity or service throughout the network, or gaining access to and from different points across the network. In general, the analysis of the shortest distances between all pairs of nodes and the distribution of the path lengths may reveal important information about the levels of efficiency and accessibility in a network and will be partially correlated with reliability in terms of meeting the objectives of the system. In WDNs, this matter is worth examining due to the benefits that a well-conceived optimally-connected network layout might bring in terms of better reachability between sources and consumers, the quality of the service provided (water quality, quantity and pressure) and the efficient management of resources by suppliers (reduction in the financial, energy and other costs associated with network design, maintenance and operation). Water distribution infrastructures are spatial networks organized against the Euclidean plane with their components occupying actual physical locations. Two basic measures to inform the level of accessibility and efficiency in such networks are Euclidean distances between nodes and geodesic path lengths.\\

Specific to WDNs is the concept of hydraulic head (i.e. conservation of energy) on which the efficient cost-effective operation of pressurized WDNs largely depends. Opposite to the notion of conservation of energy in WDNs is the concept of energy losses observed as: (i) major losses due to pipe wall frictions, and (ii) minor losses due to turbulence and changes in streamlines through fittings and junctions\cite{21}. The Darcy-Weisbach equation for pressure loss due to friction along a given length of pipe between two nodes $i$ and $j$ is given by $\Delta{p}=f\frac{L_{ij}}{D_{ij}}\frac{\rho V^2}{2}$ where $f$ is the pipe friction factor, $L_{ij}$ is the pipe length, $D_{ij}$ is the pipe diameter, $\rho$ is the density of the fluid and $V$ is the average velocity of the fluid flow. Consequently, it is observed that the hydraulic efficiency at the pipe level and across the network, among other factors, depends  on the pipe lengths, where larger diameter short pipes have smaller energy losses. The minor losses, on the other hand, occur at fittings and junctions whereas a smaller number of such fittings potentially means smaller number of medium junctions between two nodes and hence smaller path lengths. In practice, minor losses are sometimes accounted for by the equivalent pipe length method \cite{21}. With these simplifications, we relate the hydraulic efficiency of water supply systems to the patterns of distribution of Euclidean pipe lengths, and to some extent the distribution of geodesic path lengths of the underlying network.\\

\begin{figure}[t]
\centering
\includegraphics[width=0.5\textwidth, height=12cm]{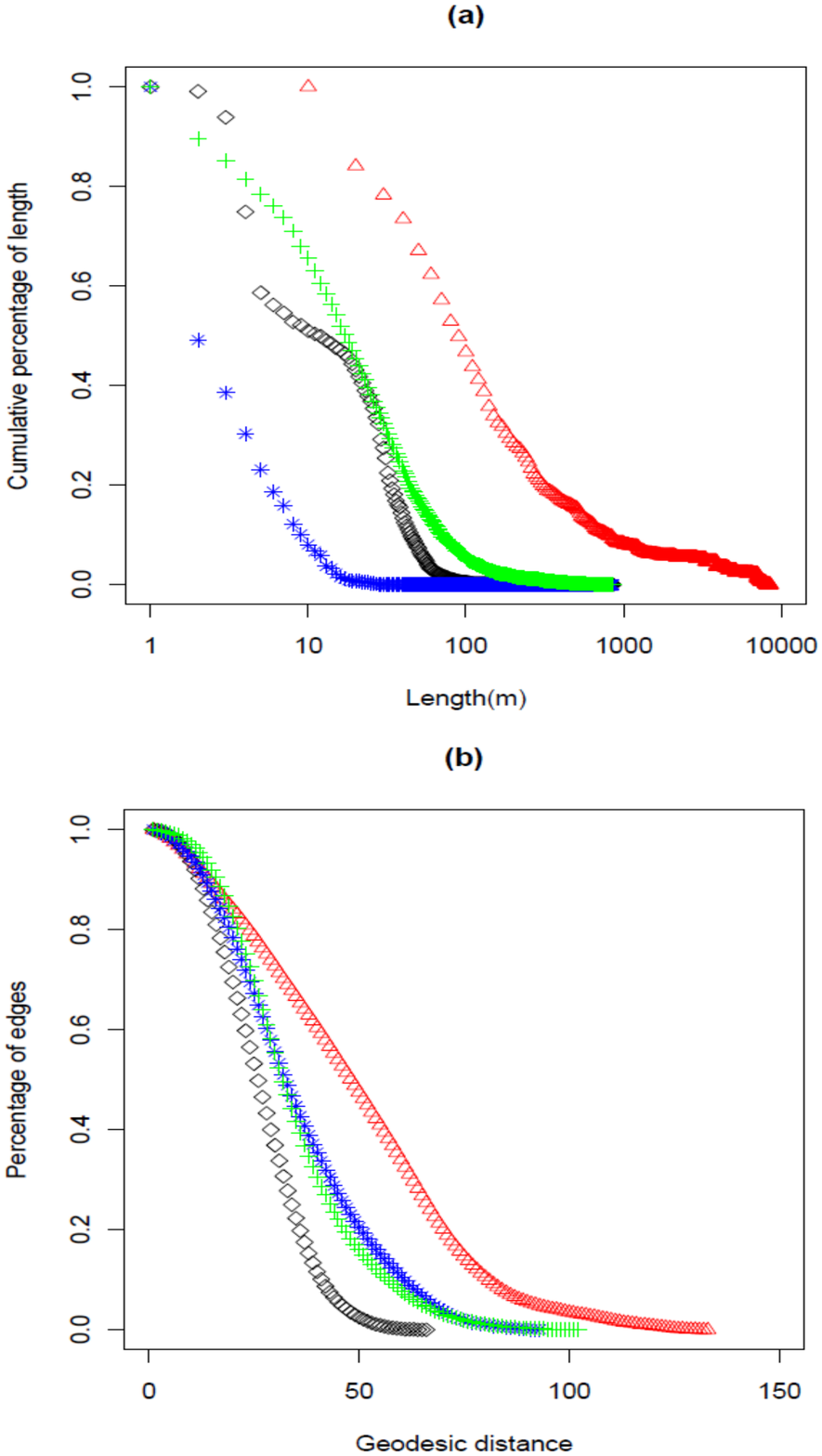}
\caption{Distribution of (a) the Euclidean edge lengths, and (b) geodesic distances for Richmond (red triangles), Colorado Springs (black diamonds), East-Mersea (blue stars) and Kumasi (green crosses).}
\end{figure}
The local structural organization of WDNs and the distribution of edge lengths can be viewed as a replica of the local urban structure, but also as an indication of the preferred actions taken by suppliers to reduce the construction and supply costs and to minimize energy losses. Therefore, in a typical WDN it is expected that short and medium size supply pipes at denser grid-like core sub-networks near the city centers or urban areas will predominate, with very few long distance pipes to carry water from the suburban sources and reservoirs. Studying the (cumulative) distribution of the edge lengths for the benchmark WDNs confirms this proposition, as illustrated in Fig. 3(a). The shortest geodesic path length $d_{ij}$ between two nodes $i$ and $j$ in an undirected connected graph is the minimum number of edges traversed in order to reach node $j$ from node $i$. Graph diameter $d$ is the maximum graph eccentricity defined as the maximum value of such shortest paths. The characteristic path length l is defined as the average of the shortest path lengths
\begin{equation}
\ l=\frac{1}{n(n-1)} \Sigma_{(i\neq j)} d_{ij}                                                                                                                                                                                            
\label{Equation 3}
\end{equation}
which represents the average degree of separation between all graph nodes. The (cumulative) distribution of the geodesic distances for the studied networks is shown in Fig. 3(b), with the values for average pipe length, mean geodesic path length and network diameters provided in Table (II). In this sense, the studied structures deviate significantly from small world networks, despite the prevalence of the short (in terms of Euclidean distances) edges, where near-planarity of the networks is perceived to be the reason \cite{41}.\\

The largest deviation from an efficient small world structure (illustrated by largest Euclidean as well as geodesic path length) has been observed in the Richmond network, probably due to geographical spread of the network combined with largely irregular non-mesh structure. Colorado Springs on the other hand is a smaller world than the others (as evidenced through a smaller graph diameter, shorter characteristic path lengths). This, viewed in line with the local meshed structure and the presence of mostly short and medium size pipes, indicates properties of more equalized distribution of pressure and flow across the system and overall operational and hydraulic efficiency.\\

One of the most important objectives in the operation of WDNs is to maintain the path connectivity between the source(s) and the consumers (network nodes) and make such path connectivity as short and efficient as possible. Therefore, instead of assessing the efficiency based on the connectivity between all pairs of nodes, it is more appropriate to measure the efficiency based on the connectivity between a root node such as a reservoir and other nodes in the network. One such measurement is known as the network's route factor \cite{42} and defined as 
\begin{equation}
\ g= \frac{1}{(n-1)} \Sigma_{i=1}^{n-1} \frac{\epsilon_{s,i}}{\delta_{s,i}}                                                                                                                                                                                                                                                                                                                                                                                    
\label{Equation 4}
\end{equation}
where $\epsilon_{s,i}$ is the combined (Euclidean) distance along the edges connecting node $i$ to the source $s$, and $\delta_{s,i}$ is the direct Euclidean distance. The smallest possible value of the route factor is 1 which is characteristic of a star graph (with all its nodes directly connected to the source), regarded as the optimal network in the sense that it has short and efficient paths to the source. A greater value of the route factor means greater deviation from the optimal network structure and hence greater costs and effort required to construct and operate the network, dispatch the utility across or navigate through different routes.\\
\begin{table}[t]
\centering
\caption{Path length and efficiency measures ($d$ = diameter, $l$ = characteristic path length, $a_l$= average pipe length(m), $g$ = route factor)}
\label{tab:2}   
\begin{ruledtabular}  
\begin{tabular}{ccccc}
\noalign{\smallskip}
Network                        &  $d$                           & $l$                                  & $a_l$                         & $g$  \\
\noalign{\smallskip}\noalign{\smallskip}
East-Mersea                 &  97                              & 34.48                             &  $27.52$                     &  1.54  \\ 
Colorado Springs          &  69                              & 25.94                             &   $187.12$                  &  1.45  \\ 
Kumasi                            &120                           &  33.89                             &   $316.20$                  &  1.46  \\ 
Richmond                        &135                           &  51.44                            &   $633.09$                   &  1.67  \\ 
\noalign{\smallskip}
\end{tabular}
\end{ruledtabular}
\end{table}
The route factor measurements for the studied networks are presented in Table (II). The range of values obtained for the route factors are similar to those reported by Gastner and Newman \cite{43} for other types of spatial distribution networks (such as $g=1.13$ for the western Australian gas pipelines and $g=1.59$ the sewer system). Authors of \cite{43} discuss that real world networks "appear to find a remarkably good compromise" between the two extreme models of star graph (optimal in the sense of having short, efficient paths to the root) and minimum spanning tree (optimal in the sense of having minimum total edge length). With the route factors close to one, studied WDNs are remarkably efficient in this sense, despite the lack of a central plan to improve global efficiency, due to contemporary planning and management strategies that seek to optimize local network robustness, reliability and efficiency as suggested elsewhere \cite{34}. In this study, the value of the route factor for networks with multiple reservoirs (i.e. multiple root nodes) has been obtained by taking the average over the individual route factors for each water supply source in the network. In the calculation of the route factors, only reservoirs and large volume water supply sources are regarded as root nodes.
\section{Robustness and structural vulnerability of water distribution networks}
\label{sec:5}
The analysis of structural vulnerability is carried out by studying the network topology and connectivity configurations and monitoring the changes in system functionality following perturbations such as single or multiple component removals as a result of either random failures or targeted attacks. This entails measuring important operational indicators such as diameter, efficiency and local or large scale connectivity for the initial network and for the network post-intervention.  The use of alternatives to connectivity measures as a way of indicating network vulnerability was first proposed by Bollob$\acute{a}$s \cite{44} in the context of communication networks. The work considered connectivity optimization problems with respect to changes in network diameter due to link or node failures. Alternatively, one could use a statistical approach to assess network robustness \cite{8} by measuring the fraction of the nodes (edges) to be removed before complete defragmentation (large scale disconnection) happens. Such measurements of network robustness have been exercised by performing random failure scenarios or attack simulations based on deletion in decreasing order of the most central or most connected nodes/links. Such studies have found that certain network topologies (e.g. scale-free networks) are extremely vulnerable to targeted attacks on their hubs \cite{8}.\\

In general, network topologies are similar to one of the groups of: centralized (e.g. wheel or star-like), decentralized (e.g. hub-spoke) or distributed structures, depending on the formation processes and organizational hierarchy of network components. Usually, centralized or hub-spoke structures provide greater operational efficiency and reliability, but also higher vulnerability to targeted attacks on their hubs \cite{8}. However, as discussed earlier, WDNs are spatially organized homogeneous networks with no pronounced hubs and generally structured in a distributed fashion. In other words, non power law random distribution of the node degrees implies that most of the nodes and links have comparable importance from the point of view of degree-based random failures or targeted attack strategies, and consequently no avalanche breakdown may happen following the removal of such components \cite{45}. \\

The threshold for random removal of nodes for any degree distribution \cite{46} is given by
\begin{equation}
\ f_c= 1- \frac{1}{\frac{<k^2>}{<k>}-1}                                                                                                                                                                                                                                                                                                                                                                                   
\label{Equation 5}
\end{equation}
which provides a theoretical value for the critical fraction of the nodes which need to be removed for a network to lose its large scale connectivity (i.e. complete destruction of the largest cluster). Using this analytical formula for the studied WDNs, it is found that: $f_{c}=0.42$ for Colorado Springs, $f_{c}=0.37$ for Kumasi, $f_{c}=0.32$ for Richmond and $f_{c}=0.22$ for East-Mersea. In other words, a complete disintegration of Colorado Springs takes the removal of about 42 percent of its nodes and the adjacent connections while, by removal of 38 percent of the nodes Kumasi will become completely disintegrated and so on. The above discussion is of some practical value only if the vulnerability is considered when WDNs are exposed to extreme events and catastrophes. However, other structural vulnerability measurements may reveal important or even more useful information on the current structure of WDNs and enable the comparative study of structural vulnerability. \\

The most important operational objective of WDNs is to supply clean water from the source to consumers with sufficient quantity and pressure and there seems to be a strong relationship between the type and the location of failures and the network's capability to meet its objectives. Whilst the operational consequences of the failure of certain WDN components may be tolerated by using redundant appurtenances and re-routing the flow, the removal of even a tiny percentage of the nodes or links (e.g. those directly connected to water supply sources) may completely disrupt network operation. Consequently, WDNs can be regarded as extremely vulnerable to the removal of certain nodes such as the reservoirs and water supply sources and their adjacent links. In other words, the hubs in WDNs should be seen as not necessarily the highly connected or the most central components; but they are rather the most influential ones (such as the source nodes and their adjacent nodes and links) or the components most critical to the satisfaction of specific network objectives (such as the nodes and links whose failures may disconnect the source from a large part of the network).\\

\begin{figure}[]
\centering
\includegraphics[width=0.5\textwidth, height=7cm]{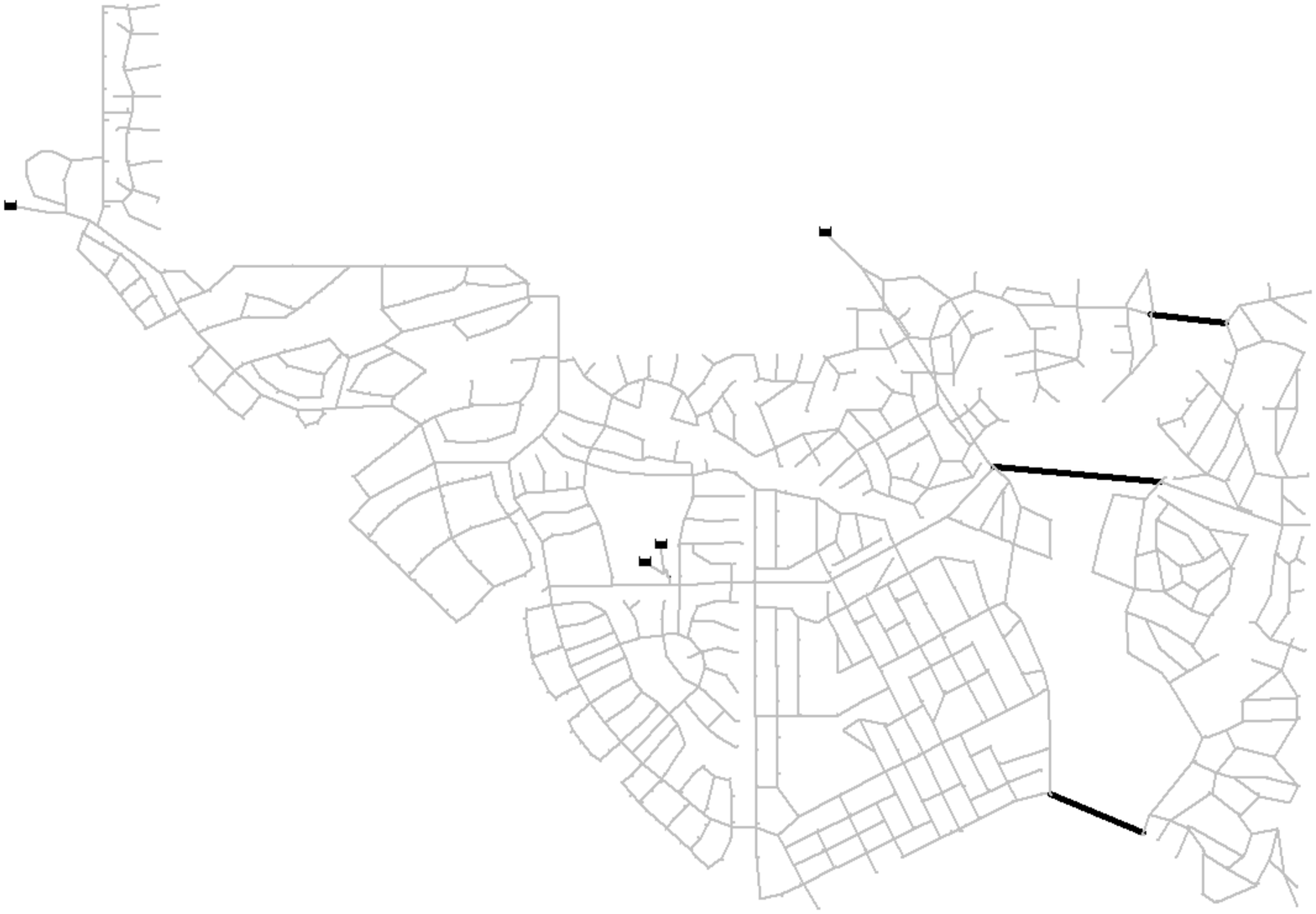}
\caption{A highlighted cut-set in the Colorado Springs network consisting of three bridges whose simultaneous removal creates disconnection between the water supply sources (black rectangles) and a large fragment of the network.}
\end{figure}
To this end, the structural vulnerability and robustness of a WDN may be investigated by quantifying the level of optimal-connectivity of network design, by identifying critical locations and the most influential components followed by studying their failure consequences on network performance. One way to identify the critical locations in WDNs is by detecting the cut-sets i.e. the sets of components whose removal results in disconnection. More specifically, a node cut-set (edge cut-set, respectively) is a set of nodes (edges, respectively) whose removal disconnects a specified pair of nodes \cite{35}. A single element node cut-set is called an articulation point and a single edge cut-set is called a bridge (Fig. 4). The node-connectivity (edge-connectivity) is the smallest number of nodes (edges) whose removal disconnects the network (or in case of disconnected networks it increases the number of connected components). These metrics can also be viewed as simple robustness indicators which quantify the minimum number of attacks or failures required to render a group of the nodes disconnected. In the studied WDNs however, these measurements become trivially equal to one, due to the sparse structure of networks and the fact that most of the end-users are supplied by single connections only. Therefore, it would be necessary to employ other measurements to differentiate between structural vulnerability and fault-tolerance of these networks.\\
      
Here, such measurement has been carried out by studying the spectrum of the network adjacency matrices of studied networks. Two such measurements utilized here are algebraic connectivity and spectral gap, network properties that quantify the robustness and optimal connectivity of sparse networks, independent from the network size or drawing. Algebraic connectivity $\lambda_2$ first introduced in \cite{47} and extensively discussed in \cite{48,49} is defined as the second smallest eigenvalue of the normalized Laplacian matrix of a network. The Laplacian matrix of $G$ with $n$ nodes is a $n \times n$ matrix $L=D-A$ where $D=diag(d_i)$ and $d_i$ is the degree of node $i$ and $A=(a_{ij})$ is the adjacency matrix of $G$ where $a_{ij}=1$ if there is a link between nodes $i$ and $j$ , and $a_{ij}=0$ otherwise. The smallest eigenvalue of a Laplacian matrix is zero with its multiplicity equal to the number of a network's connected components. Algebraic connectivity is a positive value whose magnitude indicates network robustness and well-connectedness. Larger values of algebraic connectivity represent higher robustness against efforts to decouple parts of the network. Spectral gap $\Delta$ is the difference between the first and second eigenvalues of the adjacency matrix $A$. \\

A sufficiently large value of spectral gap is regarded as a necessary condition for the so-called "good expansion" properties \cite{13} and the relative similarity between the network structure and a family of optimally-connected sparse regular networks known as "expanders" \cite{50}. The lack of good expansion, represented by small spectral gap, is usually observed through simultaneous low connectivity, sparseness and the presence of bridges and articulation points whose removal results in the split of the network into two or more large fragments. On the other hand, the existence of good expansion together with uniform degree distribution, results in higher structural robustness against node and link failures. A summary of the evaluated vulnerability measurements is presented in Table (III). It should be noted that while the obtained values for algebraic connectivity are typically very low (mainly due to the discussed planarity and lack of intents to optimize global invulnerability), they provide a useful general way of comparing network robustness against the removal of nodes and links following failures or targeted attacks.
\begin{table}[]
\centering
\caption{Spectral and global measurements for the benchmark water networks ($\lambda_2$ = algebraic connectivity, $\Delta $= spectral gap, $f_c$ = critical ratio of defragmentation)}
\label{tab:3}  
\begin{ruledtabular}    
\begin{tabular}{cccc}
\noalign{\smallskip}
Network                        &        $\lambda_2$                          &  $\Delta$                     &      $f_c$      \\
\noalign{\smallskip}\noalign{\smallskip}
East-Mersea                &         $1.97\times10^{-4}$                   & $3.91\times10^{-2}$        &  0.22            \\
Colorado Springs       &         $2.43\times10^{-4}$                   & $2.83\times10^{-2}$        &  0.42            \\
Kumasi                          &         $9.40\times10^{-5}$                   &  $9.08\times10^{-3}$       &  0.37            \\
Richmond                     &         $6.09\times10^{-5}$                   &   $7.27\times10^{-2}$     &  0.32               \\ 

\noalign{\smallskip}

\end{tabular}
\end{ruledtabular} 
\end{table}
\section{Discussion and conclusions}
\label{sec:6}
In this paper, a complex network approach was adopted to studying the structure and vulnerability of water distribution networks. Water distribution networks are viewed as complex networks represented by link node graphs of interconnected interacting components. Several measurements were undertaken to quantify the network structure and explain its relationship with the hierarchy, evolution, performance reliability and the vulnerability of these networks. The common characteristics of water distribution networks and other types of spatially organized networks were highlighted and their different features explored. A summary of the reported observations are as follows:

(I) The studied water distribution networks are sparse near-planar graphs whose structures largely resemble the surrounding urban areas supplied by the system. The ordering structure of these networks represents a gradual and usually unplanned expansion over time as a result of the urban dwelling developments. The planarity and other geographical characteristics prevent the formation of highly connected hubs and hence water distribution networks tend to be non-heterogeneous structures with typically low connectivity.

(II) Clustering and looped or grid-like structures take place at the distribution levels in the town centers and urban areas with higher population density and greater demand for water. Reliability and efficiency considerations give rise to the observation of greater link density and higher path redundancy provided by mostly short and small pipes in such places as compared to the sparse structure of network formed by long and larger pipes at the suburbs and transmission levels of the network. Network loops are mostly non-triangular which resulted in small values of clustering coefficient. The meshed-ness coefficient is found to be a better indicator of the status of network loops and cycles and hence a better descriptor of path redundancy.

(III) The formation, design and construction of WDNs are largely influenced by the cost of connections and pumping water from the sources to demand points, subject to geographical specifications. Consequently, the metric known as route factor, which is based on the Euclidean distances between the water supply source and demand nodes, is regarded as a more realistic indicator of network efficiency in addition to a surrogate measure of the construction costs, as compared to the topological measurement of efficiency. In this sense, studied water distribution networks show high efficiency similar to other types of reported distribution networks.

(IV) Network robustness and structural vulnerability were investigated by using techniques to identify the influential components and critical locations (e.g. articulation points and bridges) and quantifying the network's well-connectedness with respect to the existence of such locations, in the absence of degree-based hubs and given the sparse structure of networks. Descriptive measurements, including those derived from the spectral analysis of network connectivity and Laplacian matrices, quantified the level of structural network tolerance against failures and removal of components and enabled a basic comparison between different network designs.\\

Overall, these observations provide a framework for the study of water distribution systems and the level of similarity or difference between water distribution networks and other types of (spatial) network, in terms of their structure, organization, efficiency and vulnerability. While, as demonstrated, using mainly topological network techniques and measurements presents answers to several basic yet important questions regarding the structure and function of water distribution networks, a thorough assessment of system complexity, efficiency and vulnerability will require further information and specifications relating to the system and its operational status. \\

To this end, it is very important that a realistic assessment of the network structure, efficiency or vulnerability should avoid attempting an exclusive characterization of network structure or function by using only single (or even a few) network measurements as ultimate indicators. Moreover, analyses based on structural measurements should be accompanied by relevant heuristics and expert interpretations so that necessary modifications in assessment criteria and measurements can be considered. In this respect, the current analysis may be regarded as a demonstration that pure network measurements may set up some very useful and necessary but perhaps insufficient criteria for the analysis of structural reliability or vulnerability of water distribution systems and other similar spatially organized systems. Possible future work in this area may investigate issues such as network expansion strategies and trade-off scenarios of optimizing network connectivity as a function of construction costs and improvement in serviceability indicators.\\

We would like to thank the Leverhulme Trust for financial support as well as Anglian Water, the Ghana Water Company and the Centre for Water Systems at Exeter University for the network data. We also wish to thank two anonymous reviewers for their constructive and helpful comments on an earlier version of this manuscript.\\


%

\begin{thebibliography}{}
%
%

\bibitem{1}
D. J. Watts, S. H. Strogatz, Nature \textbf{393}, 440 (1998).

\bibitem{2}
A. L. Barab$\acute{a}$si, R. Albert, Science \textbf{286}, 509 (1999).

\bibitem{3}
R. Albert, A. L. Barab$\acute{a}$si, Rev. Mod. Phys.\textbf{74}, 47 (2002).

\bibitem{4}
M. E. J. Newman, SIAM Rev.\textbf{45}, 167 (2003).

\bibitem{5}
S. N. Dorogovtsev, J. F. F. Mendes, Adv. Phys. \textbf{51},1079 (2002).

\bibitem{6}
S. Boccaletti, V. Latora, Y. Moreno, M. Chavez, D.–U. Hwang, Phys. Rep.\textbf{424}, 175 (2006).

\bibitem{7}
S. H. Strogatz, Nature \textbf{410}, 268 (2001).

\bibitem{8}
R. Albert, H. Jeong, A. L. Barab$\acute{a}$si, Nature \textbf{406}, 378 (2000).

\bibitem{9}
V. Latora, M. Marchiori, (2003), Eur. Phys. J. B. \textbf{32}, 249 (2003).

\bibitem{10}
P. Crucitti, V. Latora, M. Marchiori, Physica A \textbf{338}, 92 (2004).


\bibitem{12}
J. Buhl, J. Gautrais, N. Reeves, R. V. Sol$\acute{e}$, S. Valverde, P. Kuntz, G. Theraulaz, Eur. Phys. J. B. \textbf{49}, 513 (2006).

\bibitem{13}
E. Estrada, Eur. Phys. J. B. \textbf{52}, 563 (2006).

\bibitem{14}
A. J. Holmgren, Risk Analysis \textbf{26}, 955 (2006).

\bibitem{15}
A. P. Masucci, D. Smith, A. Crooks, M. Batty, Eur. Phys. J. B. \textbf{71}, 259 (2009).

\bibitem{16}
M. Rosas-Casals, S. Valverde, R. Sol$\acute{e}$, Int. J. Bif. Chaos \textbf{17}, 2465-2475 (2007).

\bibitem{17}
R. Carvalho, L. Buzna, F. Bono, E. Guti$\acute{e}$rrez, W. Just, D. Arrowsmith, Phys. Rev. E, \textbf{80}, 016106 (2009).


\bibitem{19}
A. Cardillo, S. Scellato, V. Latora, S. Porta, Phys. Rev. E. \textbf{73}, art. no. 066107 (2006).

\bibitem{20}
A. Ostfeld, J. Water Resour. Plan. Manage. \textbf{131}, 58 (2005).

\bibitem{21}
Haestad Methods, T. M. Walski, D. V. Chase, D. A. Savic, W. Grayman, S. Beckwith, E. Koelle, \textit{Advanced Water Distribution Modeling and Management} (HAESTAD Press, 2003).

\bibitem{22}
A. Ostfeld, J. Hydroinformatics \textbf{6}, 281 (2004).

\bibitem{23}
E. Alperovits, U. Shamir, Water Resour. Res. \textbf{13(6)}, 885 (1977).

\bibitem{24}
N. Duan, L. W. Mays, K. E. Lansey, J. Hydraul. Engng. \textbf{116(2)}, 249 (1990).

\bibitem{25}
P. Jacobs, I. C., Goulter, Engng. Optim. \textbf{15}, 71 (1989).

\bibitem{26}
J. M. Wagner, U. Shamir, D. H. Marks, J. Water Resour. Plan. Manage. \textbf{114}, 276-294 (1988).

\bibitem{27}
G. C. Dandy, A. R. Simpson, L. J. Murphy,  Water Resour. Res. \textbf{32(2)}, 449 (1996).

\bibitem{28}
D. A. Savic, G. A. Walters, J. Water Resour. Plan. Manage. \textbf{123(2)}, 67 (1997).

\bibitem{29}
H. R. Maier, A. R. Simpson, A. C. Zecchin, W. K. Foong, K. Y. Phang, H. Y. Seah, C. L. Tan, J. Water Resour. Plan. Manage. \textbf{129(3)}, 200 (2003).

\bibitem{30}
I. C. Goulter, Civil Engng. Syst. \textbf{4(4)}, 175-184 (1987).

\bibitem{31}
Y. Shen, K. Vairavamoorthy, in ASCE proceedings of the International Conference on Computing in Civil Engineering, (2005), 1333-1339.

\bibitem{32}
I. Lippai, in ASCE Proceedings of the Pipeline Division Specialty Conference, (2005), p.1058.

\bibitem{33}
J. E. Van Zyl, D. A. Savic, G. A. Walters, J. Water Resour. Plan. Manage. \textbf{130},  160 (2004).

\bibitem{34}
D-H. Kim, A. E. Motter, J. Phys. A: Math. Theor. \textbf{41}, 224019 (2008).

\bibitem{35}
M. E. J. Newman, \textit{Networks, an introduction} (Oxford University Press, NY, 2010).

\bibitem{36}
L. C. Freeman, Sociometry \textbf{40}, 35 (1977).

\bibitem{37}
A. Yazdani, P. Jeffrey, in Water Distribution System Analysis Conference, WDSA2010, AZ, USA (2010).

\bibitem{38}
B. E. Larock, R. W. Jeppson, G. Z. Watters, \textit{Hydraulics of Pipeline Systems} (CRC Press LLC, Boca Raton, 2000).

\bibitem{39}
M. Barthelemy, A. Flammini, Phys. Rev. Lett. \textbf{100}, art no. 138702 (2008).

\bibitem{40}
A. Pastor-Satorras, A. Vazquez, A. Vespignani, Phys. Rev. Lett. \textbf{87}, 258701 (2001).

\bibitem{41}
M. T. Gastner, M. E. J. Newman, Eur. Phys. J. B. \textbf{49}, 247 (2006).

\bibitem{42}
W. R. Black, \textit{Transportation: A Geographical Analysis} (Guildford Press, NY, 2003).

\bibitem{43}
M. T. Gastner, M. E. J. Newman, J. Stat. Mech. P01015 (2006).

\bibitem{44}
B. Bollob$\acute{a}$s, in Theory of Graphs (G. Katona and P. Erdos, Eds.), Akad. Kiado, Budapest, 29 (1968).

\bibitem{45}
P. Holme, B. J. Kim, C. N. Yoon, S. K. Han, Phys. Rev. E \textbf{65}, 056109 (2002).

\bibitem{46}
R. Cohen, K. Erez, D. Ben-Avraham, S. Havlin, Phys. Rev. Lett. \textbf{85}, 4626 (2000).

\bibitem{47}
M. Fiedler, Czechoslovak Mathematical Journal \textbf{23}, 298 (1973).

\bibitem{48}
B. Mohar, Graph Theory, Combinatorics and Applications \textbf{2}, 871 (1991).

\bibitem{49}
Jamakovic, A., Uhlig, S., in Proceedings of the 3rd EURO-NGI Conference on Next Generation Internet Network, Trondheim, Norway, (2007), p.96.

\bibitem{50}
L. Donetti, F. Neri, M. A. Munoz, J. Stat. Mech. \textbf{8}, P08007 (2006).


\end{thebibliography}
%


\end{document}